\documentclass[useAMS,usenatbib]{mn2e}
\usepackage{epsfig}
\usepackage{amssymb}
\usepackage{amsmath}

\newcommand\ion[2]{#1\,{\scshape{#2}}}

\title[Partial obscuration of AGNs by outflowing dusty clouds] {Partial dust obscuration in active galactic nuclei as a cause of broad-line profile and lag variability, and apparent accretion disc inhomogeneities}
\author[C. M. Gaskell and P. Z. Harrington]{C. Martin Gaskell\thanks{E-mail:
mgaskell@ucsc.edu} and Peter Z. Harrington\thanks{E-mail: pharring@ucsc.edu}\\
\\Department of Astronomy and Astrophysics, University of California, Santa Cruz, CA 95064
}

\begin{document}

\date{}

\pagerange{\pageref{firstpage}--\pageref{lastpage}} \pubyear{2017}

\maketitle

\label{firstpage}

\begin{abstract} 
The profiles of the broad emission lines of active galactic nuclei (AGNs) and the time delays in their response to changes in the ionizing continuum (``lags") give information about the structure and kinematics of the inner regions of AGNs. Line profiles are also our main way of estimating the masses of the supermassive black holes (SMBHs). However, the profiles often show ill-understood, asymmetric structure and velocity-dependent lags vary with time.  Here we show that partial obscuration of the broad-line region (BLR) by outflowing, compact, dusty clumps produces asymmetries and velocity-dependent lags similar to those observed. Our model explains previously inexplicable changes in the ratios of the hydrogen lines with time and velocity, the lack of correlation of changes in line profiles with variability of the central engine, the velocity dependence of lags, and the change of lags with time. We propose that changes on timescales longer than the light-crossing time do not come from dynamical changes in the BLR, but are a natural result of the effect of outflowing dusty clumps driven by radiation pressure acting on the dust. The motion of these clumps offers an explanation of long-term changes in polarization. The effects of the dust complicate the study of the structure and kinematics of the BLR and the search for sub-parsec SMBH binaries. Partial obscuration of the accretion disc can also provide the local fluctuations in luminosity that can explain sizes deduced from microlensing.
\end{abstract}


\section{Introduction} 

An important, ubiquitous feature of thermal active galactic nuclei (AGNs)\footnote{see \citealt{Antonucci12} for a discussion of thermal versus non-thermal AGNs.} is the presence of broad emission lines with widths of up to many thousands of km sec$^{-1}$. The gas producing these lines is called the broad-line region (BLR).  Any successful model of the workings of the inner regions of AGNs must explain the existence of the BLR and the broad line profiles.  Bright, blue, optically-selected AGNs, such as the vast majority of AGNs in the SDSS, show roughly symmetric, centrally-peaked BLR line profiles (see, for example, \citealt{Baldwin75}).  However, there are also thermal AGNs selected by their extended radio emission which have less bright optical emission from the AGN compared with their host galaxies.  Because the host galaxies are more obvious these have historically been called ``radio galaxies". Their broad line profiles are often much more complex than those of bright optically-selected AGNs \citep{Osterbrock+75}.  Many theories have been proposed to explain the diversity of line profiles.  These include various geometries (spherical, disk-like, or bi-conical) and different kinematics such as random virialized motions, turbulent Keplerian discs, or outflows.

Having a correct understanding of the structure and kinematics of the BLR is important for understanding how AGNs work because
the BLR provides our main way of getting the masses of the supermassive black holes in AGNs \citep{Dibai77}. These mass estimates obviously depend on the structure and kinematics of the gas.  The low-ionization gas produces strong Balmer lines.  These can show double-peaked line profiles that \citet{Oke87} argued were best explained by a disc origin.  \citet{Alloin+88} also argued that the broad Balmer lines of the optically-selected, radio-quiet AGN Akn 120 arise from a disc.  Subsequently, \citet{Chen+89} showed that the broad H$\alpha$ profile of the radio galaxy Arp 102B was particularly well fit by a disc model.  The consensus now is that emission from the low-ionization BLR gas comes predominantly from a rotating, flattened configuration of turbulent gas (\citealt{Shields77,Shields78,Osterbrock78}; see \citealt{Gaskell09} for a review).  The more symmetric, centrally-peaked profiles of bright AGNs are also explained by the same distribution of gas, but viewed close to face-on (see, for example, Figure 2 of \citealt{Gaskell11}).

The flux of broad lines responds to changes in the ionizing continuum on a light-crossing time (typically light-days to light-months) \citep{Lyutyi+Cherepashchuk72}.  This gives constraints on the size and structure of the BLR and other reprocessing regions.  Following \citet{Blandford+McKee82} this is commonly called ``reverberation mapping".  The most widely used approach has been to cross-correlate continuum light curves with line-flux light curves \citep{Gaskell+Sparke86,Gaskell+Peterson87} to determine an average time delay (the ``lag") and get an estimate of the size of the reprocessing region.  Furthermore, for simple symmetric models of the BLR, determining this delay as a function of {\em velocity} gives the net radial velocity of the BLR \citep{Gaskell88}.  This is because when an emission line responds to a change in continuum level we see the material closest to our line of sight respond first and then material on the far side. Velocity-resolved reverberation mapping of many objects over the past 30 years (see \citealt{Gaskell+Goosmann13} for a review) confirm that, on average, the motions of both the low- and high-ionization BLR gas are virialized with a slight net inflow.  The results of reverberation mapping and line profile studies thus support our picture of the low-ionization BLR being in a flattened, rotating, turbulent distribution.

The main problem with the turbulent Keplerian disc model is that while broad Balmer line profiles, or the profiles of at least
the variable part of the lines, {\em sometimes} show the predicted double-peaked profile (e.g., \citealt{Chen+89} or \citealt{Shapovalova+04}), these cases are the exception.  The more usual situation is for the line profiles to be asymmetric (see the compilation by \citealt{Eracleous+Halpern94,Eracleous+Halpern03}) with either the blue or red displaced peak being stronger.  Various ideas have been offered to explain these asymmetries.  Models include binary supermassive black holes \citep{Gaskell83}, conical outflows \citep{Zheng+91}, asymmetric discs \citep{Eracleous+Halpern94}, asymmetric distributions of BLR clouds \citep{Wanders+Peterson96}, warped discs \citep{Wu+08}, and off-axis flares \citep{Gaskell10b,Gaskell11}.

A further problem for explaining line profiles is that not only do line profiles rarely match the predicted double-peaked profile expected from an inclined disc, but {\em the profiles change}.  These changes do not show any clear correlation with  changes in the overall continuum level \citep{Wanders+Peterson96}.  Also, changes in total line intensities are frequently not explained by changes in the continuum level \citep{Shapovalova+10}.

There are problems with interpreting reverberation mapping results too.  Analysis of reverberation-mapping data from the multi-year NGC~5548 campaigns showed that the derived transfer functions (i.e., the impulse responses - see \citealt{Blandford+McKee82}) for different observing seasons differed by more than the experimental errors (see Figure 8 of \citealt{Pijpers+Wanders94}). Also, the lag time between line and continuum variability (the first moment of the transfer function) shows significant changes {\em over an observing season} (see Figure 1 of \citealt{Maoz94}).  The transfer function is normally interpreted as giving information about the distribution of the responding material (in this case the BLR). This material cannot redistribute on a timescale faster than the orbital timescale (many years or decades).

Another problem raised by reverberation-mapping observations is that while velocity-resolved time delays in general favour virialized motion with a slight net inflow \citep{Gaskell+Goosmann16}, there have been a couple of notable exceptions.  \citet{Kollatschny+Dietrich96} found that the blue side of the \ion{C}{iv} profile of NGC~5548 temporarily led the red side (see their Figure 7).  In a symmetric BLR scenario this implies a strong outflow.  However, as \citet{Gaskell+Goosmann13} point out, in this case, the blue side leading certainly did {\em not} result from a real outflow because because the analysis of \citet{Kollatschny+Dietrich96} also shows that only three months later the {\em red} side of \ion{C}{iv} led, which would imply inflow.  Another apparent exception to the general picture of a gravitationally-bound BLR was NGC~3227 in 2007, for which \citet{Denney+09} found variability of the blue side of H$\beta$ leading variability of the red side.  They interpreted this as an outflow signature.  \citet{Gaskell10b} showed that velocity-dependent lags in NGC~3227 and other AGNs could be explained simply by an asymmetry in the lags on either side of the BLR disc.

In this paper we propose a new explanation of the changes in line profiles and in the velocity-dependent lags: {\em the motion of small, dusty clouds across our line of sight to the BLR.}  We discuss our motivation for this in Section 2.  In Section 3 we describe our modelling of line profile changes, and in Section 4 we describe our modelling of the velocity dependence of lags. In Section 5 we discuss how modest patchy obscuration will mimic the accretion disc inhomogeneties proposed to explain micro-lensing results.  We discuss various implications of our model for spectroscopic searches for close supermassive binary black holes, explaining changing-look AGNs, modelling of the BLR, interpreting reverberation mapping results, and understanding the dust geometry.

\section{Small absorbing clouds along the line of sight to AGNs} 

\subsection{Evidence for the existence of small clouds}

Intrinsic absorption by gas along the line of sight to AGNs has been known from the earliest days of AGN studies.  The
first line detected in an AGN \citep{Mayall34} was the \ion{He}{I}$^* \lambda$3889 line in NGC~4151.  This line is interesting because it implies that the absorbing gas has a high density and column density (see \citealt{Liu+15}). \citet{Anderson+Kraft71} showed that the velocity structure in the \ion{He}{I}$^* \lambda$3889 absorption was variable on a timescale of only a few years and \citet{Anderson74} argued that the gas was located at a distance of about one light-month from the center and had an electron density $n_e > 10^{6}$ cm$^{-3}$. \citet{Hamann+95} found UV absorption line variability in the luminous, high-redshift AGN UM~675 -- an AGN about a 1000 times more luminous than NGC~4151 and hence with a BLR about 30 times larger.  \citet{Hamann+95} argued that if the change in UM~675 was due to the transverse component of velocity of the absorbing gas cloud, the size of the cloud is of the order of 0.01 pc (i.e., a couple of light weeks). This makes it smaller in size than the BLR of UM~675.

From studies of absorption-line profiles in Q0059-2735 \citet{Wampler+95} found that ``condensations produce . . . saturated lines that fail to occult completely the central continuum sources or the low-ionization emission-line sources" and also that  ``while the low-ionization BAL gas occults the nuclear continuum source, it fails to occult the low-ionization broad
emission line region."  Absorption line profile variability and partial covering (which are now known to be common -- see, for example, \citealt{Barlow+97}) thus both point to the existence of absorbing clouds that are small compared with the size of the BLR.  For example, \citet{Shi+16} find that the variable, blueshifted, high-velocity Balmer line absorption in SDSS J125942.80 + 121312.6 is well explained by changes in the covering factor due to transverse motion of the absorbing clouds.

\citet{Barr+77} discovered that the X-ray absorption in NGC~4151 showed strong variability.  They suggested, following Anderson \& Kraft, that this was due to clouds moving across the central X-ray and optical emitting regions.  \citet{Holt+80} showed that the X-ray spectral shape and variability was consistent with partial coverage by clouds that were smaller than the continuum-emitting region.  Observations of additional objects \citep{Reichert+85} showed that partial coverage was common.  There have now been many reports of occultation events in X-ray light curves (see \citealt{Turner+Miller09} and \citealt{Markowitz+14} for reviews.)  For example, \citet{Lamer+03} observed an occultation event in 2000/2001 lasting about a year for NGC~3227 where the absorbing column density rose from around zero up to $2.5 \times 10^{23}$ and back to around zero.  Observations of another event in 2008 \citep{Beuchert+15} indicated that the transiting cloud had an irregular density profile. X-ray absorption depends on the total column density and degree of ionization and will be seen regardless of whether dust is present or not.  \citet{Kaastra+14} present detailed X-ray and UV spectroscopy of NGC~5548 showing the effects of a clumpy stream blocking 90\% of the soft x-ray emission and causing simultaneous deep, broad UV absorption troughs. The outflow
velocities are a few 1000 km s$^{-1}$ and they estimate the distance to be only a few light days from the black hole.
The above-mentioned transit events seen in X-rays and the UV can be of dust-free gas.  There is evidence, however, for a significant component of X-ray absorbing gas with dust mixed in with it -- the so-called ``dusty warm absorber" (e.g., \citealt{Mathur+94}, \citealt{Brandt+96}).  This shows that dust can survive in the ionized gas of the ``warm absorber".

Many AGNs show broad absorption line (BAL) systems.
Broad absorption line quasars (BALQSOs) show strong X-ray absorption (\citealt{Mathur+95}; see \citealt{Fan+09} for more extensive observations). \citet{Weymann+91} noticed that the UV continua of BALQSOs in general, and LoBALs in particular, were redder than non-BALQSOs. \citet{Sprayberry+Foltz92} showed that these differences were consistent with increased reddening. From an analysis of spectra of over 10,000 BALQSOs, \citet{Gaskell+16} have shown that the mean {\em additional} reddening of AGNs with BALs is $E(B-V) \sim 0.03 - 0.05$ and that the dust has a steep, SMC-like reddening curve.  Putting these results together, we see that both X-ray observations and UV/optical absorption line studies clearly point to the inner regions of AGNs often being partially covered by clouds with sizes comparable to the size of the accretion disc and BLR.  Furthermore, there is evidence that {\em at least some of this gas is dusty.}  This is not a rare phenomenon. The fraction of AGNs BALs is high. After allowing for selection effects \citet{Dai+08} find that 44\% of AGNs at any epoch have BALs. Around redshift 2 the fraction of rises to $\sim 80$\% (see Figure 27 of \citealt{Allen+11}).

\subsection{Intrinsic hydrogen line ratios}

The strongest broad optical lines, and the ones in which profile irregularities are most readily observed, are the Balmer lines.  Simple recombination theory predicts that the flux ratio of H$\alpha$/H$\beta$ (which we will call ``the Balmer decrement") should be the so-called ``Case B" value of $\sim 2.7$ (see \citealt{Osterbrock+Ferland06}).  However, observations of the integrated flux of broad lines almost always give Balmer decrements that are steeper than Case B.  Also, the observed ratio of Lyman $\alpha$ to the Balmer lines is almost always less than the Case B ratio.  These problems were referred to as ``the hydrogen-line problem" and considerable effort was spent over several decades in trying to solve this problem by invoking radiative transfer effects and collisional enhancement (see \citealt{Gaskell17} for a detailed review).  These interpretations have been widely accepted.  In this picture the hydrogen line ratios will depend on the physical conditions and radiation field and are expected to change with conditions.

An alternative but less popular solution to the hydrogen-line problem was proposed by \citet{Shuder+MacAlpine77,Shuder+MacAlpine79}. They suggested that the observed non-Case B ratios were due to {\em reddening}.  \citet{Dong+05,Dong+08} showed by considering blue AGNs that were presumed to have low reddening, that, contrary to the predictions of models invoking radiative transfer and collisional effects, there was little or no variation in the intrinsic Balmer decrement. Further analysis by \citet{Gaskell17} indicated that the blue AGNs of \citet{Dong+08} were somewhat reddened and that after allowance for this the intrinsic H$\alpha$/H$\beta$ ratio was a Case B value.  Analysis of the Lyman $\alpha$/H$\beta$ ratio showed that it too was a Case B value.  The hydrogen line ratios can thus be used as a reddening indicator and steeper Balmer decrements are due to dust.  It has also been proposed that when there are changes in Balmer decrements these are simply due to changes in the reddening \citep{Goodrich89,Goodrich95}. \citet{Wang+09} present good evidence for this happening in Mrk 1393 where there was an associated change in the X-ray absorption.

\subsection{Velocity-dependent Balmer decrements}

Attempts to explain hydrogen line ratios have mostly focussed on ratios of total line intensities.  There has been much monitoring of broad-line variability in AGNs for reverberation mapping campaigns, but these observations have largely focussed on just the H$\beta$ line because it is well placed for observation even at moderate redshifts and the nearby [\ion{O}{iii}] $\lambda$5007 line provides a convenient relative flux calibration of continuum variability.

The object with the best published observations to date of the profiles of {\em both} H$\alpha$ and H$\beta$ is NGC~4151
for which \citep{Shapovalova+10} present 10 years of monitoring of the Balmer line profiles. Their Figures 14-17 show significant changes occurring in the Balmer decrement both with time and {\em velocity}. Their observations of the H$\alpha$/H$\beta$ ratio as a function of time and velocity are reproduced here in Figure 1.

It can be seen from Figure 1 that there is a clear velocity dependence in the Balmer decrement changing over a time scale of
six months or so. Since the Balmer lines arise predominantly from a flattened, rotating distribution of gas (see \citealt{Gaskell09}), the velocity field is ordered and a velocity dependence of the H$\alpha$/H$\beta$ ratio means that {\em the ratio changes with position in the BLR.}

If the ratio is governed by radiative transfer and collisional effects, then the ratio should depend on the radiation field.  \citet{Shapovalova+04,Shapovalova+10} show that the total H$\alpha$/H$\beta$ ratio does {\em not} show a simple dependence on the flux level, and it is hard to see how small localized regions of the flattened BLR producing the Balmer lines at particular velocities could experience very different physical conditions, especially on timescales observed.

If instead the Balmer decrement is an indicator of reddening, this implies that the reddening is also velocity-dependent.  We show here how small, dusty clouds crossing the BLR can indeed produce both line profiles and temporal changes in the velocity-dependent Balmer decrement similar to those observed.  The gas emitting the Balmer lines produces a characteristic double-peaked profile when the BLR is viewed off-axis, as is more likely to be the case when objects are radio selected. Therefore, when double peaks are observed, we know the AGN is not being viewed exactly face on. We propose that in these AGNs, outflowing dust clouds can obscure parts of the BLR such that extinction of the broad-line emitting gas occurs only over a limited range of velocity.

\begin{figure}
\centering
\includegraphics[width=8.5cm]{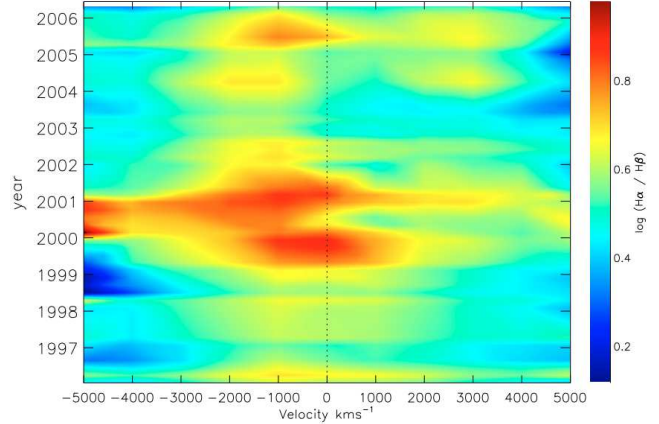}
 \caption{The H$\alpha$/H$\beta$ ratio as a function of velocity and time for NGC 4151. The bar on the right shows the colour corresponding to the logarithm of the H$\alpha$/H$\beta$ ratio.  Turquoise corresponds to a Case B value and redder colours indicate  higher reddening.  Note that the observed ratio is uncertain at high velocities ($|v_{\rm rad}| > 4000$ km/s)
 because of difficulties in setting the continua levels and blending by other lines (see Fig. 7 of \citealt{Shapovalova+10}). Figure reproduced by permission from \citet{Shapovalova+10}. This figure is available in colour in the on-line version of the paper.}
\end{figure}

\section{Modelling line profiles} 

A computer code, \texttt{BL-RESP}, has been developed to model BLR observations (see \citealt{Gaskell10b,Gaskell11}).  \texttt{BL-RESP} generates many observable quantities and makes movies of BLR cloud motions seen from various viewing angles.  Outputs include profiles of emission lines, and the temporal response of the BLR flux as a function of time delay or lag,  $\tau$, to continuum variability (normally expressed by the transfer function, $\Psi(\tau)$; \citealt{Blandford+McKee82}). \texttt{BL-RESP} does {\em not} include detailed atomic physics.  For additional discussion of the code, see \citet{Gaskell10b}.
To match the stratified distribution of gas in the GKN model (see \citealt{Gaskell09}), the BLR structure is approximated as a spherical distribution with polar bi-cones removed.  The flattened turbulent disc model is reproduced by setting a large opening angle for the bi-cones. A real BLR probably consists of a quasi-continuous fractal distribution of clouds \citep{bottorff+ferland01}, but for ease of computation \texttt{BL-RESP} assumes that the BLR consists of condensations (``clouds'').

Because pressure is inadequate to support the observed thickness of the BLR there must be vertical motion of the gas.  This is also required to explain line profiles and the consistency of mass estimates from the BLR.  It is likely that, as in the solar chromosphere, magnetic effects are responsible for the vertical motions.  However, because there is not yet a satisfactory theory of BLR motions, \texttt{BL-RESP} simply models the vertical structure and motions by having clouds move in tilted circular orbits.  This gives simple harmonic motion in the vertical direction.  This is also effectively simultaneously modelling the case where clouds are thrown up from the accretion disk and fall back down again onto the disk with behavior similar to flares on the sun.  The latter model avoids the problem of how the clouds survive a hypersonic passage through the accretion disk.  This model assumes that BLR clouds are effectively collisionless and that their motions are dominated by gravity.  This has the advantage of involving a known force.
The distribution of tilts is taken to be uniform. Model parameters are the inner and outer radii, the radial distribution of clouds, the viewing angle, the distance dependence of the response to the continuum, and the magnitude of a possible radial component of the cloud velocity.

In this paper we are only considering Balmer lines. The relative range of radii over which the Balmer lines are emitted is well-determined observationally (a) from the widths of $\Psi(\tau)$ recovered from reverberation mapping observations and (b) from line profile fitting.  Reverberation mapping can be used to set the scale factor.
For H$\beta$, the range of radii over which the line is emitted is surprisingly narrow.  The best H$\beta$ transfer functions give a range of a factor of four or so between the inner and outer radii (see Fig.~8 of \citealt{Pijpers+Wanders94}).
Such a range of radius is consistent with inner and outer radii deduced by \citet{Eracleous+Halpern03} from fitting disk-like line profiles.

The response of BLR emission at a given location to continuum variability depends on several factors including: the ionizing flux, $F_{ion}$, reaching the gas, the amount of gas at the location (which depends on the gas density and local filling factor), and details of photoionization and recombination.  The ionizing flux reaching the gas will fall off with the distance the radiation has to travel, $r$, at least as fast as $r^{-2}$.  It will fall off faster if absorption and scattering along the line of sight are significant.  In \texttt{BL-RESP} the fall-off in $F_{ion}$ is parameterized as $F_{ion} \propto r^{-\alpha}$.  For the models in this paper the index, $\alpha$, was taken to be $2$.  For symmetric illumination the effect of this distance dependence of $F_{ion}$ is coupled with the effect of radial fall off in surface density of the gas, which we took to be $r^{-1}$.  This gives an overall emissivity power-law index, $q=3$.  This is the approximate value suggested by disk profile fitting by \citet{Eracleous+Halpern03} (see their Table 4).  The results presented here are not sensitive to the precise values of $\alpha$ and $q$.

Spatially-resolved {\it Hubble Space Telescope} observations of the narrow-line region (NLR) of NGC~4151 \citep{Fischer+13} show that NGC~4151 is inclined at $45^{\circ}$ to our line of sight.  \citet{Marin16} has shown that orientations derived from the NLR are well correlated with other, more indirect indicators of orientation on smaller scales.  We therefore take the inclination of the BLR in our models to be $45^{\circ}$. BLR cloud motions are assumed to be dominated by gravity.  We took the thickness of the distribution of BLR clouds producing the Balmer lines to be $\sim \pm 20^{\circ}$ from the mid-plane.  This angle sets the ratio of the vertical (``turbulent") component of velocity to the dominant Keplerian component of velocity. These parameters give line profiles similar to those of \citet{Eracleous+Halpern94,Eracleous+Halpern03}. We do not consider absorption by gas in our model because Balmer lines arise from excited states.

\subsection{The effect of small opaque dust clouds crossing the BLR}

In Figure 2 we show the resulting line profile when a small optically-thick cloud passes in front of a BLR.
It can be seen that only part of the line profile is changed. The strongest effect on the line profile is when the opaque dust cloud passes in front of the inner region of the BLR with high-velocity gas and the highest
surface brightness. This was calculated by removing the contribution to the line profile from the occulted region of the disk.

\begin{figure}
\centering
\includegraphics[width=8.6cm]{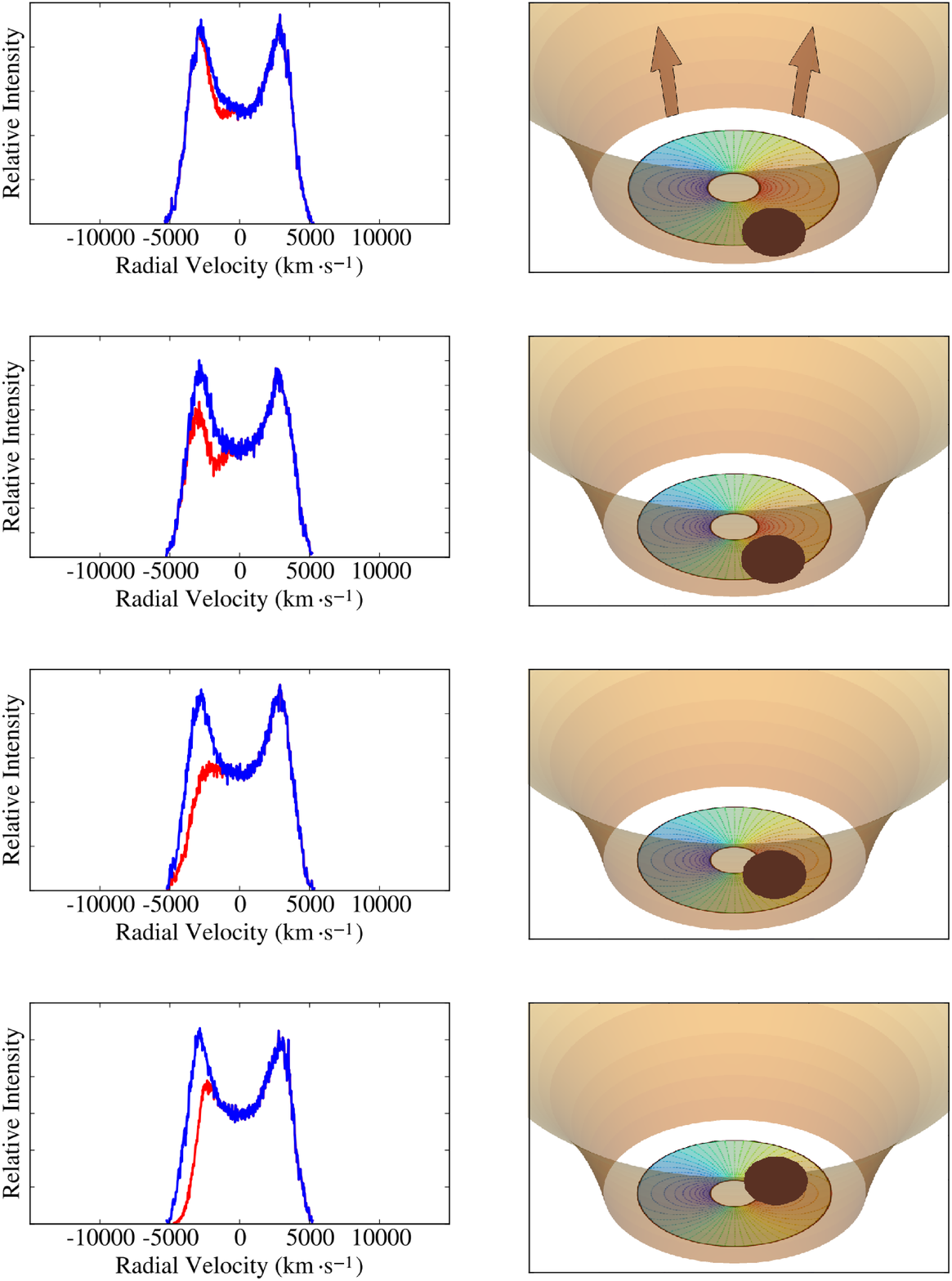}
 \caption{The effect of an optically-thick cloud of arbitrary elliptical shape obscuring part of the Balmer-line emitting BLR. The BLR is taken to be tilted at an angle of $45^{\circ}$. The different coloured regions of the BLR (not to scale) indicate the shape of iso-velocity contours of the BLR. Since the dust clouds are driven outwards by radiation pressure, they will flow along a hyperbolic cylinder (shown in light brown, arrows indicate the direction of outflow). The location of the obscuring cloud on the surface of the hyperbolic cylinder is shown by the dark brown oval. The unobscured line profile is shown in blue in the corresponding panels on the left and the obscured profiles are shown in red. This figure is available in colour in the on-line version of the paper.}
\end{figure}

The first result that can be seen from Figure 2 is that a relatively small dust cloud, covering less than 10\% of the BLR,  has a significant effect on the line profile.  Furthermore, the resulting profiles are similar to observed ones (see, for example, the atlas of profiles in \citealt{Eracleous+Halpern03}).

\subsection{Modelling temporal changes in the velocity dependence of the Balmer decrement}

If a dust cloud has a finite optical thickness (for example, an optical depth in the $V$ band, $\tau_V$, of, say, 2, which corresponds to $E(B-V)\sim 0.75$) there will be greater extinction at shorter wavelengths.
This will steepen the Balmer decrement over the range of velocity affected by the cloud. As shown in Figure 1, velocity-dependent Balmer decrements are observed. The dusty clouds causing this cannot have high optical depths.  We now demonstrate how dust clouds of finite optical depth partially obscuring the BLR can reproduce observed changes in the Balmer decrement with velocity and time similar to those shown in Figure 1.

If it be true, as the evidence indicates, that higher H$\alpha$/H$\beta$ ratios are simply due to reddening, then temporal variability of the ratio must be due to {\em changes} in the reddening.  This is consistent with the finding of \citet{Heard+Gaskell16} that much of the dust reddening AGNs is located close to the centre, between the narrow-line region (NLR) and the BLR. Since the BLR size is on the order of light days to light weeks for typical, well-studied, nearby AGNs, and velocities in the outer BLR are $>1000$ km s$^{-1}$, variability of the extinction over periods of months to years is possible. We demonstrate here how the observed changes in H$\alpha$/H$\beta$ ratios with velocity and time naturally arise if the dust
is in a clumpy outflow originating on the edge of the BLR.

Radiation pressure on dust drives powerful outflows in stars.  Reverberation mapping shows that the dust sublimation radius of AGNs is just beyond the BLR, in agreement with theory \citep{Suganuma+06}. Radiation pressure will expel dusty clouds at the outer edge of the BLR. They will partially obscure the BLR and accretion disk when the system is not viewed close to face-on.

It can be seen from Figure 1 that the observed timescale for changes in H$\alpha$/H$\beta$ in NGC~4151 is on the order of 9 months. During the period of monitoring by  \citet{Shapovalova+10}, H$\beta$ lagged the continuum by 11 +/- 4.5 days.  The less certain lag of H$\alpha$ of 22 +/- 10 days is consistent with this. We will therefore adopt an effective size of 11 light-days for the Balmer line emitting region.  If we associate the timescale of Balmer decrement variability with the time for a dust cloud to cross the BLR, it implies a transverse velocity on the order of 10,000 km/sec.  This is typical of outflow velocities of transient broad absorption line systems (e.g., \citealt{Leighly+15}), but greater than the orbital velocity at the edge of the BLR.

By moving dust clouds of finite optical depth across the BLR in \texttt{BL-RESP} we are able to qualitatively reproduce temporal variations in the velocity dependence of the Balmer decrement similar to those observed.  We give two examples in Figure 3 of an oval dust cloud with $E(B-V) \sim 0.75$, elongated in the vertical direction, moving upwards across the BLR.  The locations of the dusty clouds are shown in panel c.  Panel a shows the change over time of the velocity dependence of H$\alpha$/H$\beta$ for a cloud moving across the inner part of the BLR.  This part of the BLR does not produce much Balmer line emission. Panel a shows how a steepening of the Balmer decrement over a narrow range in velocity, such as happened in 1999 or 2005 (see Figure 1), can occur. The precise wavelength of the center of the change in Balmer decrement depends on how close the cloud is to the axis.  Panel b shows how simultaneous variation of both high- and low-velocity Balmer decrements (such as was observed in late 2001) is produced when the moving cloud occults the inner region of the BLR producing the high-velocity wings of the Balmer lines. Although we have not tried to match the observations in detail, it can be seen that we can qualitatively reproduce the changes observed. The Balmer decrement averaged over the entire line profile for the position producing the largest effect on the line profile is 2.80 for cloud a (corresponding to panel a), and 2.83 for cloud b (corresponding to panel b). Thus, the effect on the integrated H$\alpha$/H$\beta$ ratio is not large and the main effect is on the line profile. Because the effects of small dust clouds on the integrated Balmer decrement is small, partial obscuration of the sort modeled here will not strongly affect the observed correlation between the integrated H$\alpha$/H$\beta$ flux ratio and the continuum level in NGC 4151 (see, for example, \citealt{Rakic+17}). When there is a correlation between the integrated Balmer decrement and the continuum, this would be due to larger dust clouds.

\begin{figure}
 \centering \includegraphics[width=8.5cm]{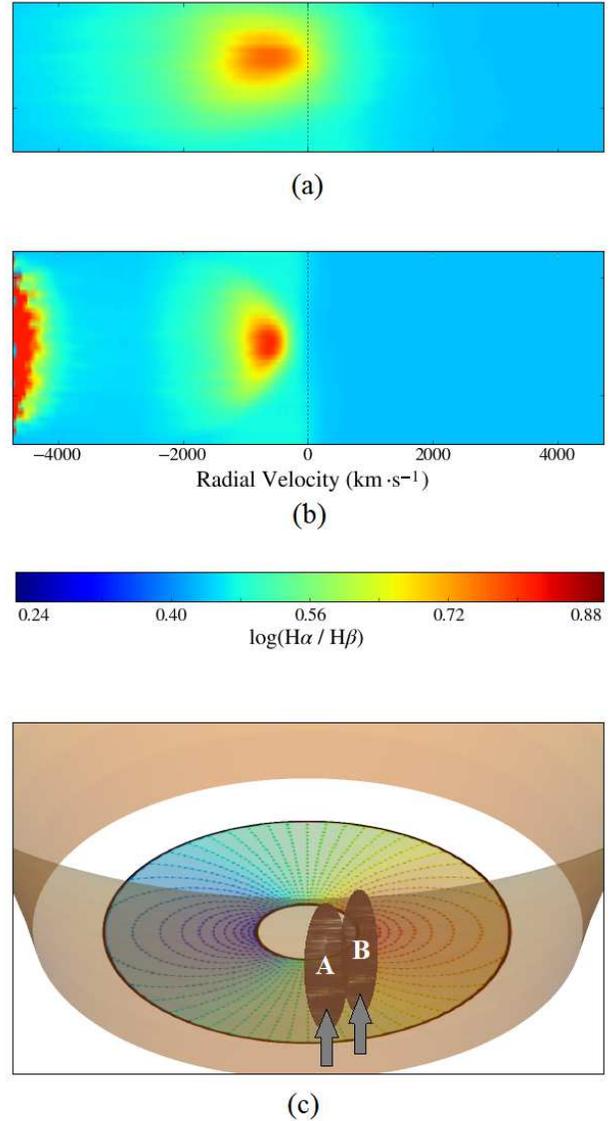}
 \caption{
 Panels (a) and (b) on the top show simulated H$\alpha$/H$\beta$ ratios as a function of velocity and time for optically-thin
 clouds passing in front of a BLR modeled as shown in panel (c). The cloud position labelled ``A" corresponds to panel a, and  ``B" corresponds to panel b.  The amount of reddening ($E(B-V) \sim 0.75$) has been chosen to approximate the  observed Balmer decrements in Figure 1.   In panel a the cloud is mostly moving upwards across the hole in the middle of  the BLR (i.e., the region not producing much Balmer line emission). In panel b the cloud clips the high-velocity inner edge of the Balmer line emitting region as it moves upwards.  This figure is available in colour in the on-line version of the paper.}
\end{figure}

\section{Modelling Velocity-Dependent Lags} 

\subsection{Changing transfer functions}

As noted above (see Introduction), the changes in the NGC~5548 transfer function cannot be due to physical changes in the geometry of the BLR.  However, patchy obscuration will distort the transfer function.  For example, if there were a dust cloud blocking the near side of an inclined, flattened BLR this would suppress the transfer function at short time delays and make the overall lag be longer.  This is illustrated in the cartoon in Figure 4.  Conversely, if the small dust cloud preferentially blocks the far side, the lag will be reduced.

\begin{figure}
 \centering \includegraphics[width=8.5cm]{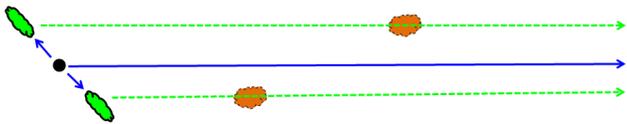}
 \caption{ Cartoon illustrating why a small dust cloud will strongly affect the observed lag.  Ionizing radiation from the central regions near the black hole (solid blue lines) comes directly to the observer and is also reprocessed in the flattened BLR (assumed to be tilted at $\sim 45^{\circ}$).  The reprocessed radiation (e.g., H$\beta$), shown by the dashed green lines, also travels to the observer.  Because of the light-travel-time delay, the line emission from the near side of the  BLR arrives first and the light from the far side somewhat later. When there is a dust cloud on the line of sight to the near side of the BLR this attenuates light from this side and results in a net increase in the lag.  Likewise, a cloud obscuring the far side decreases the lag.}
\end{figure}

\subsection{Apparent changes in direction of BLR motion in velocity-resolved transfer functions}

As discussed in the Introduction, differences in lags between the blue and red sides of lines have been taken as an indicator of the amount of radial motion.  However, as indicated in Figure 4, a small dust cloud can change a lag.  If this happens on only one side of the BLR disc, as shown in Figure 2 and 3, this will cause the variability of one side of the line to lag the other.  Such a difference mimics a signature of radial motion \citep{Gaskell10b}.

In Figure 5 we show the velocity-dependent lags derived by \citet{Pei+17} for H$\beta$ in NGC~5548 from observations in the first and second halves of the 2014 monitoring campaign.  The dashed blue line in Figure 5 shows the velocity-dependent lags for an unobscured BLR.  This has been calculated assuming the same BLR parameters described above, an inclination to our line of sight of $30^{\circ}$, and a half-opening angle of $85^{\circ}$.  The size scale (i.e., the mean lag) has been chosen to match the average lags from the entire observing period and the velocity scale has been chosen to match the line width.  The observed lags are obviously quite different for the two halves of the observing campaign and each is different from the prediction shown by the dashed blue line.

The disc-like BLR has a very structured velocity field, as shown by the iso-velocity contours in the top half of Figure 6; there is also distinct, but different, structure in the lags across a disc-like BLR, as shown by the iso-delay contours in the bottom half of Figure 6. Since the iso-velocity and iso-delay contours are roughly orthogonal over much of the BLR, if changes in velocity-dependent lags are due to obscuration by dust clouds, the coupling of the velocity and lag constrains where the obscuring clouds have to be located. A detailed fitting to the observations is beyond the scope of this paper but we find that just two dust clumps passing in front of the BLR during each half of the observations can fit the lags.  The resulting lags are shown by the red lines in Figure 5.  These models have the same BLR parameters and velocity/lag scales as the unobscured model given by the blue lines.  It can be seen that, depending on the location of the clumps, the lag is either increased or decreased.

\begin{figure}
 \centering \includegraphics[width=9cm]{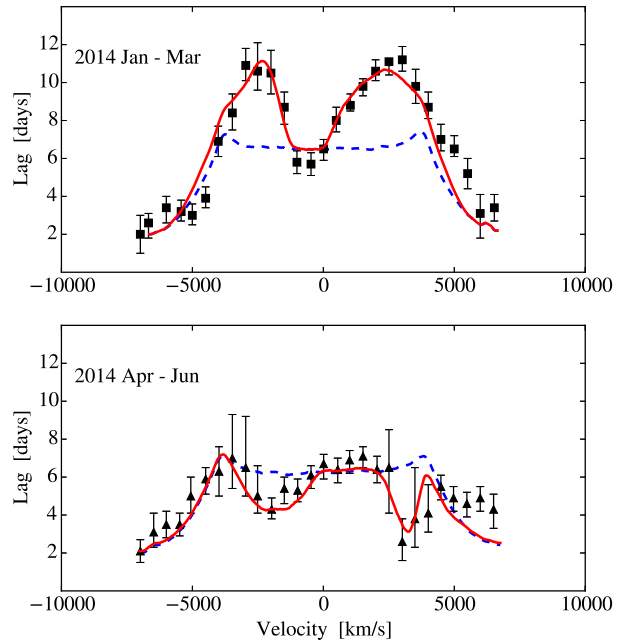}
 \caption{Observed lags of H$\beta$ in NGC~5548 relative to the continuum as a function of velocity for the first and second halves of the 2014 campaign \citep{Pei+17}. The red lines show our derived velocity-dependent lags resulting from dust concentrations in locations as shown in Figure 6 below.  The dashed blue lines show the velocity-dependent lags for no obscuration or uniform obscuration. This figure is available in colour in the on-line version of the paper.}
\end{figure}

\begin{figure}
 \centering \includegraphics[width=6cm]{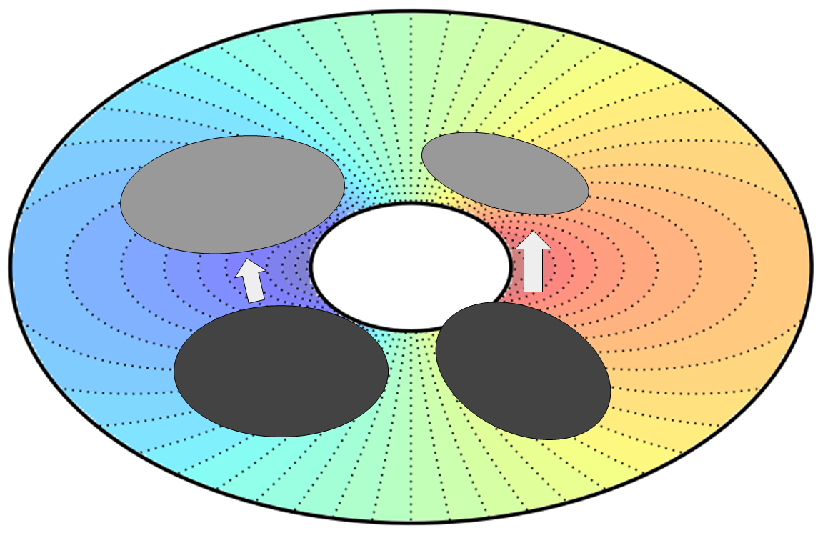}
 \centering \includegraphics[width=6cm]{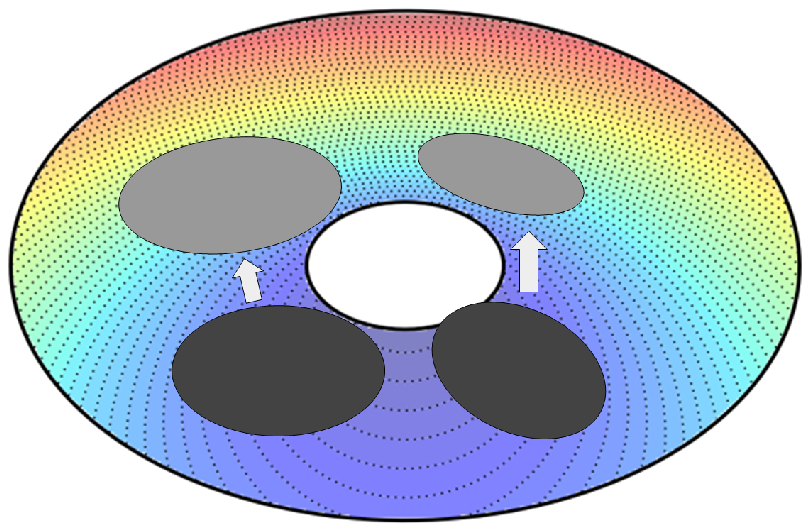}
 \caption{Deduced positions of the dust clouds blocking the BLR during the first quarter of 2014 (black ovals) and the second quarter (grey ovals). These are the time periods of the observations giving the upper and lower halves of Figure 5. The top image shows relative iso-velocity contours for an idealized flat Keplerian disk inclined at $45^{\circ}$ to our line of sight and the lower image shows iso-delay contours on the same disc. This figure is available in colour in the on-line version of the paper.}
\end{figure}

\section{Apparent accretion disc inhomogeneities} 

Standard accretion disc theory predicts that temperature, $T$, in an accretion disc varies with radius as $R^{-3/4}$ \citep{Lynden-Bell69}. Microlensing of AGNs gives an indication of the size of an accretion disc as a function of wavelength, $\lambda$.  Studies of this have revealed two inconsistencies with simple accretion disc theory (see \citealt{Jimenez-Vicente+14} and references therein).  The first is that the sizes of discs are systematically larger than predicted from the observed fluxes.  This size discrepancy is removed when the observed flux is corrected for extinction \citep{Gaskell17}.  The second problem is that microlensing implies steeper temperature gradients that the standard theory.  The eight AGNs studied by \citet{Jimenez-Vicente+14} give an effective temperature gradient of $T \propto R^{-1.3 \pm 0.25}$.

\citet{Gaskell08,Gaskell10b} argues that the emission from an accretion disc is probably not axisymmetric.  \citet{Dexter+Agol11} proposed that accretion disc inhomogeneities can explain the microlensing results. As they explain, the local inhomogeneous spectra peak at a lower flux than a single temperature blackbody. Thus, to produce the same total flux at a given wavelength, the
emission must arise from a larger area. Therefore the disc appears larger at that wavelength (see their Figure 2).    \citet{Gaskell08,Gaskell10b}, \citet{Dexter+Agol11} and \citet{Ruan+14} suggest that the inhomogeneities are intrinsic to the disc (i.e., some regions are hotter than others).

 Clumpy obscuration, caused by the transits of small dust clouds, of the {\em accretion disc} (rather than the Balmer-line emitting BLR, which has been modeled in the preceding sections) is another factor that can cause apparent inhomogeneities in the emission from accretion discs. We illustrate this in Figure 7 where we show the effect of two arbitrary clouds.

\begin{figure}
 \centering \includegraphics[width=9cm]{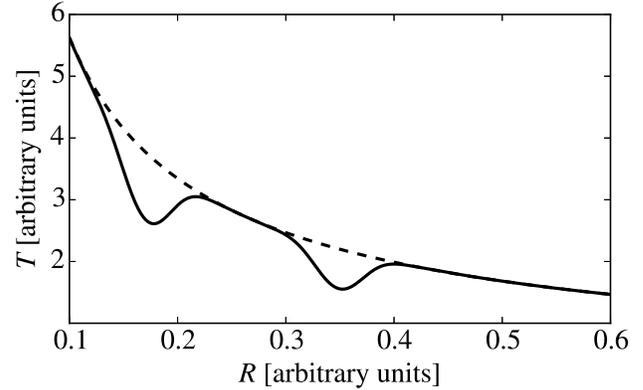}
 \caption{Illustration of the effect on the deduced  apparent radial temperature structure of an accretion disc when there is partial obscuration by two small dust clouds at different radii.}
\end{figure}

\citet{Dexter+Agol11} could increase the size of the accretion disc found by microlensing by invoking fluctuations in the luminosity of about 15\% (see their Figure 2).  This can readily be explained in the optical by invoking local fluctations in $E(B-V)$ of only 0.05.

\citet{Dexter+Agol11} did not consider the problem of the steep temperature gradient found by microlensing \citep{Jimenez-Vicente+14} since this was not recognized at the time of their paper.  However, this can also naturally be explained both in their model with temperature fluctuations in the disc and by patchy dust.  This is because fluctuations in $T$ cause bigger fluctuations in flux at shorter wavelengths, and fluctuations in extinction also cause larger fluctuations in flux at shorter wavelengths.

\section{Discussion} 

If our model is correct, the effect of partial occultation by dust complicates the determination of the structure of the BLR. We mention here some things that will be affected and compare the predictions of our model and other models.

 \subsection{Spectroscopic searches for supermassive binaries}

Some AGNs have broad Balmer line profiles showing peaks that are highly blueshifted or redshifted.  It has been
proposed that these are the result of supermassive binary black holes \citep{Gaskell83}, each with an associated BLR. There has been considerable interest in recent years in searching for close (sub-parsec) supermassive binaries using broad line profiles (see \citealt{Popovic12} for a review).  However, profile variability on light-crossing timescales is inconsistent with the binary black hole model and favours the displaced peaks arising in a rotating disc \citep{Gaskell10a}.  Since the classic double-humped profile is not seen, we are not seeing axially-symmetric emission. Such profiles can instead be explained by warped discs \citep{Wu+08} or non-axisymmetric continuum emission \citep{Jovanovic+10,Gaskell10b,Gaskell11}, but partial occultation of the BLR by dust, as we are proposing here, offers a natural way of explaining displaced BLR peaks as well (see Figure 2).  We will demonstrate in a separate paper (Harrington \& Gaskell, in preparation) that partial obscuration can readily explain the extremely asymmetric profiles such as those suggested by \citet{Tsalmantza+11} and \citet{Eracleous+12} as candidate supermassive binaries. We have two predictions for AGNs with extremely asymmetric profiles.  The first is that the depressed sides of the line profiles (regardless of whether they are on the blue or red side) should have a Balmer decrement that is steeper on average than the other side.  Our second prediction is that objects with extremely asymmetric profiles should eventually revert back to looking like more typical AGNs.

Displaced peaks in AGNs show slow changes in radial velocity and these changes have been interpreted as evidence for orbital motion \citep{Gaskell96,Bon+12,Bon+16,Li+16}.  Figure 2 demonstrates, however, that such radial-velocity changes can be produced by obscuring clouds crossing our line of sight to the BLR.  The change in velocity of the blue peak in Figure 2 is qualitatively similar to the observed velocity change of the blue peak of 3C~390.3 shown in \citet{Gaskell96}.  We suggest that variable, velocity-dependent extinction is the cause of the apparent velocity shifts of components of Balmer lines in well-studied AGNs such as NGC~4151 \citep{Bon+12} and NGC~5548 \citep{Li+16,Bon+16}. Clearly the presence of variable velocity-dependent extinction complicates the spectroscopic search for supermassive binary black holes.  Since the dust will also effect the continuum brightness, this could explain why line profiles and the continuum flux can show similar quasi periodicities such as those observed in NGC~4151 and NGC~5548 (see \citealt{Bon+12} and \citealt{Bon+16}).

It is hard to conceive of a way in which the relatively rapid changes in the velocity-dependent lags (see Figure 5) might be explained by having two continuum sources, each associated with one of the black holes in a supermassive binary.  For example, the changes in Figure 5 cannot be explained by linear superpositions of two separate, physically-reasonable transfer functions for two BLRs around two black holes. The largest changes are at intermediate velocities of approaching and receding gas.

\subsection{``Changing-look" AGNs}

Some AGNs, referred to as ``changing-look" AGNs, show dramatic changes of Seyfert type with the BLR either disappearing or appearing.  The changes could be due to either major changes in the accretion rate or to changes in the dust obscuration or a combination of both (see discussion in \citealt{Oknyansky+17}).  If the changing-look phenomenon is due to changing dust geometry then we predict that these AGNs are very likely to show variable asymmetric humps in their Balmer line profiles.  \citet{Oknyansky+17} report just such a change in the changing-look AGN NGC~2617. NGC 4151 itself has been a changing-look AGN (\citealt{Lyutyi+84}, \citealt{Penston+Perez84}, \citealt{Oknyansky+91}).

In Figures 2 and 3 we have considered changes in the dust geometry caused by motions of the clouds.  However, another possibility is destruction of dust when the continuum level increases. \citet{Oknyansky+17} have proposed that changing-look AGNs going from type 1.8 to type 1 could be caused by sublimation of dust when the continuum level increases.  It is now well established that the radius of emission of the hottest dust in NGC~4151 varies with the activity of the central engine (\citealt{Oknyansky+99} -- see \citealt{Kishimoto+13} and \citealt{Oknyansky+14} for more recent observations).

\subsection{Inhomogeneities in the BLR gas?}

\citet{Zheng+91} proposed that a large ($\sim90^{\circ}$) patch of enhanced emission, or ``hot-spot", within the flattened disc component of the BLR producing most of the Balmer line emission could account for changes in the displaced H$\alpha$ peaks of 3C 390.3. A similar model was explored for Arp 102B by \citet{Newman+97}.  However, \citet{Gezari+07} find that modelling profile changes by hot spots gives discrepant black hole masses for different epochs. \citet{Gilbert+99} suggest that regions of enhanced emission could be due to changing structure in the BLR or warping of the BLR disc.  However, they note that there is a timescale problem.  The proposed hot spot of \citet{Newman+97} has an orbital period of only 2 years and only lasted for a couple of orbits. As \citet{Pronik+Sergeev06} point out, such changes are taking place on timescales much shorter than the BLR dynamical timescale and cannot be caused by redistribution of the line-emitting gas.

The presence of hot (or cold) spots within the BLR could explain some of the velocity-dependent changes in the lags, because they would enhance (or diminish) emission only within a region of specific velocity and lag as discussed above.  However,
changes in velocity-dependent lags (see Figure 5) occur on timescales shorter than the BLR dynamical timescale.  They thus cannot be caused by inhomogeneities within the BLR.

A further problem for hot spots is that  emission-line theory gives no clear predictions of how the Balmer decrement would change with inhomogeneities in the BLR, while this arises naturally with changing extinction.

\subsection{Implications for reverberation mapping campaigns}

Our model has practical implications for reverberation mapping campaigns.  It needs to be recognized that observed continuum variability is not just due to intrinsic variability of the inner accretion disc, but can also be due to changing extinction. In planning future reverberation mapping campaigns it is important to observe {\em both} H$\alpha$ and H$\beta$ in order to be able to study changes in extinction. If changes in line profiles are due in part to partial obscuration by dust this also offers an explanation of why the long-term continuum level and line profiles show some connection as is seen in both NGC~4151 \citep{Bon+12} and NGC~5548 \citep{Bon+16}.

Variability of the observed flux because of changing extinction will introduce scatter in the relationship between the radius of the BLR and the continuum level.  This can be tested by looking at the Balmer decrement.  We predict that on time periods much longer that the light crossing time there will be a correlation between the Balmer decrement and the continuum level.  We do {\em not} predict that there will be a correlation between the Balmer decrement and the continuum level on short timescales (only a few times the light-crossing timescales).  We propose that in analyzing reverberation mapping data the first step should be to use the Balmer decrement to correct for changing extinction.  If available, the near IR flux will provide a useful check since it responds to the total bolometric luminosity.  On the other hand, for studying long-term, secular changes in line profiles, it is necessary to take out the effects of short-term changes in the intrinsic continuum level.  Note that the changes in the velocity-dependent Balmer decrement shown in Figure 1 are all on a timescale much longer than the light-crossing timescale.

The responses of line ratios on different timescales are an important test of our model.  If changes in the hydrogen line ratios are due to changes in the microphysics (i.e., changes in ionization parameter, density, and optical depth) driven by the ionizing continuum then we expect rapid changes in the line ratios following the continuum.  If the changes in the ratios are due to changing extinction, then there will be no such changes.  As is predicted by theory and as is observed both from reverberation mapping and microlensing studies, the region emitting the continuum is about an order of magnitude smaller than that emitting the Balmer lines (a light day or so for NGC 4151 versus a couple of light weeks).  Therefore we do not generally expect correlations between profile variability and continuum variability since the dust blocking the Balmer line emitting regions would be blocking IR emitting parts of the accretion disc.  We would also not expect a correlation between continuum variability and line profiles changes because the continuum is intrinsically very variable.  We therefore cannot make specific predictions.

\subsection{Dust geometry}

Dust will have no difficulty surviving in clouds only a few BLR radii from the center of the AGN because reverberation mapping and IR interferometry show that the radius of the emission from hot dust in NGC~4151 is only $\sim 30 - 90$ light-days (see \citealt{Kishimoto+13}, \citealt{Oknyansky+14} and references therein). Many dust geometries have been considered in the literature.  These include a simple uniform toroidal absorber, a geometrically thick torus supported by radiation pressure or turbulence, a warped or tilted disc, a clumpy torus, a dusty outflow, and nuclear starburst discs (see \citealt{Gohil+Ballantyne17} for a summary and references). In this paper we have assumed that the dust clouds crossing the BLR are part of a radiatively driven, bi-conical outflow of dust and gas (e.g., \citealt{Konigl+Kartje94}). For observational support for such a geometry in a nearby AGN see \citet{Stalevski+17}. The geometry is similar to that of the NLR found by \citet{Fischer+13} where the NLR clouds lie inside bi-cones (see many figures in their paper).  Such a model differs from the widely-considered ``clumpy torus'' dust geometry because in a bi-conical outflow, the obscuring clouds are not bound to the central AGN. Radiation pressure is a couple of orders of magnitude greater on a dust + gas mixture than on gas alone (e.g., \citealt{Scoville+Norman95}), and propels the outflow of obscuring clouds.

The effect on BLR emission line profiles is the same regardless of the obscuring cloud's origin -- it does not matter if the dusty cloud is part of an outflow {\em or} clumpy torus as long as it crosses the BLR and causes extinction over a specific range of velocity.  However, the change in line profile over time depends on the trajectory of the cloud.  Thus, given sufficiently high quality observations of H$\alpha$ and H$\beta$ over long enough time, it should be possible to distinguish between different cloud dynamical models and also to investigate the destruction of dust.  With such observations of enough objects it would be possible to investigate connections with other properties of the AGN, such as orientation and accretion rate. Unfortunately with existing technology it is not possible to resolve these clouds by direct imaging, although microlensing might give some information.  Light scattered by the dust clouds will be polarized, so motion of the dust clouds could be a cause of observed long-term changes in scattered polarized light (see \citealt{Gaskell+12}).

Absorption by {\em gas} in NGC~4151 has long been studied (see, for example, \citealt{Crenshaw+Kraemer07} and references therein).  The amount of dust needed in our model to explain the observed changes in Balmer decrements (see Figure 3) is of the order of $E(B-V) \sim 0.75$.  If we assume a standard Galactic gas-to-dust ratio $n_H/E(B-V) = 5.8 \times 10^{21}$ \citep{Bohlin+78} this gives a total column density of $4.3 \times 10^{21}$ cm$^{-2}$.  This is within the range of column density \citet{Crenshaw+Kraemer07} give based on UV and X-ray absorption (see their Table 1).  The extinction needed to explain Balmer line profile changes is thus consistent with independent estimates of the gas column densities.

\section{Conclusions}

We have reviewed and discussed multiple lines of evidence from optical, UV, and X-ray absorption studies pointing to the existence of relatively-small, dusty, rapidly-moving clouds close to the centres of AGNs and to these clouds partially occulting the continuum and BLR.  We have demonstrated how the partial obscuration by these clouds can reproduce observed Balmer line profiles, the observed changes in the H$\alpha$/H$\beta$ ratio with velocity and time, and the changing velocity dependence of time lags.  Partial obscuration of the accretion disc can also provide the local fluctuations in luminosity that can explain accretion disc sizes deduced from microlensing.  Variable, velocity-dependent obscuration complicates the spectroscopic search for supermassive black hole binaries and creates difficulties in learning about the BLR from line profiles and reverberation mapping.


\section*{Acknowledgments}
We are grateful to Ski Antonucci, Edi Bon, Xiao-Bo Dong, Wen-Juan Liu, Bill Mathews, and Victor Oknyansky for extensive comments on the paper.  We also wish to thank the anonymous referee for numerous helpful comments that have improved the paper.


\end{document}